\documentclass[]{interact}

\usepackage{epstopdf}
\usepackage[caption=false]{subfig}

\usepackage[numbers,sort&compress]{natbib}

\bibpunct[, ]{[}{]}{,}{n}{,}{,}

\theoremstyle{plain}
\newtheorem{theorem}{Theorem}[section]
\newtheorem{lemma}[theorem]{Lemma}

\newtheorem{repeattheorem}[theorem]{Axiom}

\theoremstyle{definition}
\newtheorem{definition}[theorem]{Definition}

\theoremstyle{remark}

\begin{document}


\title{A modified axiomatic foundation of the analytic hierarchy process}

\author{
\name{Fang Liu\textsuperscript{a}\thanks{Corresponding author. f\_liu@gxu.edu.cn; fang272@126.com}, Wei-Guo Zhang\textsuperscript{b}}
\affil{\textsuperscript{a}School of Mathematics and Information Science, Guangxi University, \\Nanning Guangxi 530004, China; 
\textsuperscript{b}School of Business Administration, South China University of Technology, Guangzhou, Guangdong 510641, China}
}

\maketitle

\begin{abstract}
This paper reports a modified axiomatic foundation of the analytic hierarchy process (AHP), where the reciprocal property of paired comparisons is broken. The novel concept of reciprocal symmetry breaking is proposed to characterize the considered situation without reciprocal property. It is found that the uncertainty experienced by the decision maker can be naturally incorporated into the modified axioms. Some results are derived from the new axioms involving the new concept of approximate consistency and the method of eliciting priorities. The phenomenon of ranking reversal is revisited from a theoretical viewpoint under the modified axiomatic foundation. The situations without ranking reversal are addressed and called ranking equilibrium. The likelihood of ranking reversal is captured by introducing a possibility degree index based on the Kendall's coefficient of concordance. The modified axioms and the derived facts form a novel operational basis of the AHP choice model under some uncertainty. The observations reveal that a more flexible expression of decision information could be accepted as compared to the judgments with reciprocal property.
\end{abstract}


\section{Introduction}

Since the Analytic Hierarchy Process (AHP) was developed as a choice model by \cite{Saaty1977,Saaty80}, it has been more than forty years. As shown in the AHP model, the decomposition principle is applied to construct a hierarchy of criteria, subcriteria and alternatives when faced with a complex decision making problem. The comparison technique of paired alternatives is used to construct a series of pairwise comparison matrices. The priorities of alternatives are elicited from the obtained matrices by a synthesis mechanism; then the optimal solution is reached. One can see that the investigations involving the theory and applications of the AHP choice model are never stopped \citep{Genest1996,Saaty98,Froman2001,Altuzarra2010,Bernasconi2010,Saaty2011,Saaty2013,LiuF2020}.

\begin{figure}[t]
\begin{center}
\includegraphics[height=2in]{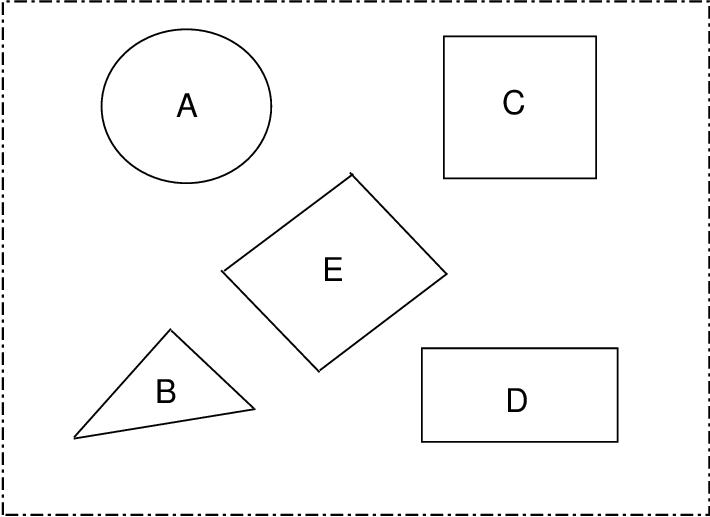}
\caption{Five figures with different areas.} \label{figure1}
\end{center}
\end{figure}

It is worth noting that one of the important issues is the axiomatic foundation of the AHP model;
and the existing one contains four axioms \citep{Saaty86}. The basic one of the four axioms is the reciprocal property of pairwise comparisons.
For convenience, we assume that there is a finite set of alternatives $\mathbb{X}=\{x_{1},x_{2},\cdots,x_{n}\}$ in a decision-making problem. The reciprocal property means that when the relative importance of $x_{i}$ over $x_{j}$ is determined as $a_{ij},$ the relative importance of $x_{j}$ over $x_{i}$ is automatically obtained as $a_{ji}=1/a_{ij}$ for $i,j\in \mathbb{I}=\{1,2,\cdots,n\}$ \citep{Saaty1977,Genest1996,Bernasconi2010}. It is obvious that the reciprocal relation $a_{ji}=1/a_{ij}$ is based on the mathematical intuition \citep{Harker1987}. We want to ask whether the relation $a_{ji}=1/a_{ij}$ is really satisfied, when the comparison ratios $a_{ij}$ and $a_{ji}$ in relative measurement are determined by evaluating the preference intensity of the alternatives $x_{i}$ over $x_{j}$ and $x_{j}$ over $x_{i},$ respectively.
For example, as shown in \cite{Saaty2013}, we compare the areas of the five figures in Figure \ref{figure1} by eyeballing them. If the reciprocal property is not assumed in advance,
it seems difficult to obtain the reciprocal relation of pairwise comparisons. The underlying reason could be attributed to the fact that various cognitive distortions of the decision maker could affect the evaluation of her/his ratio judgements \citep{Narens1996,Luce2002,Bernasconi2008}. One can find that although the reciprocal property is in agreement with the mathematical intuition, it may be not always satisfied in a practical case. In other words, the reciprocal property is a strict mathematical relation under an ideal case, and there is some deviation from the flexible expression of human-originated information. Motivated by the above consideration, we attempt to analyze the case without the reciprocal property to form a novel axiomatic foundation of the AHP model. The concept of reciprocal symmetry breaking is introduced to capture the flexible case without reciprocal property such that the uncertainty experienced by the decision maker can be naturally coped with.

Moreover, it is noted that the AHP model has also incurred some criticisms \citep{Belton1983,Belton1985,Dyer1990a,Dyer1990b,Smith2004}.
One of the criticisms comes with the phenomenon of ranking reversal. This means that when adding or deleting an alternative and/or a criterion, the ranking of the old or the remaining alternatives could be changed. That is, the ranking of alternatives is sensitive to the actions of adding and/or deleting an alternative and/or a criterion, even if the comparison ratios are all consistent.
The AHP model was criticized as an arbitrary method and it should be corrected by synthesizing the concepts of multi-attribute utility theory or the others \citep{Dyer1990a,Dyer1990b}.
The defenders considered that the phenomenon of ranking reversal is acceptable in a practical case and exhibits the structural dependence \citep{Saaty1984,Harker1987,Harker1990,Saaty1990}.
A further comment showed that the situation of applying the AHP method should be identified by comparing the multi-attribute utility theory and carrying out an example in the field of voting \citep{Perez1995}. It seems that there does not exist a widely accepted conclusion for the serious controversy \citep{Winkler1990,Gass2005}.
Fortunately, it has shown that the decision-making procedure based on the AHP model is reliable according to the recent results \citep{Bernasconi2010}. Moreover, from the existing discussions about the ranking reversal phenomenon, numerical examples are always carried out and it is a lack of a formally theoretical analysis except for the finding by \cite{Saaty1984b}. In this study, we revisit the phenomenon of ranking reversal through a theoretical investigation under the modified axiomatic foundation. The concept of ranking equilibrium is introduced to characterize the situations without ranking reversal. The likelihood of ranking reversal is quantified by proposing a possibility degree index such that the decision maker can be reminded for the occurrence of ranking reversal in terms of a percentage.

The structure of the paper is organized as follows. In section 2, the axiomatic foundations of the AHP model are focused on. The novelty comes with the modified axioms and the concept of
reciprocal symmetry breaking of pairwise comparisons. Section 3 gives some interesting results derived from the modified axioms. Some important issues are addressed such as the properties of reciprocal symmetry breaking, the approximate consistency of pairwise comparison matrices and the priorities of alternatives.
Section 4 offers a theoretical analysis for the phenomenon of ranking reversal. The novel concept of ranking equilibrium is proposed to describe the ideal cases without ranking reversal. The breaking degree of ranking equilibrium is captured by introducing a measurement index. In section 5, a practical case of decision making is restudied and some comparisons are offered to show the difference from the typical AHP model. The main conclusions are covered in Section 6.

\section{Axiomatic foundations of the AHP model}

In this section, we first recall the existing axioms of the AHP model proposed by \cite{Saaty86}. Then it is an attempt to modify the basic one such that some uncertainty of human-originated information can be captured naturally.
It is considered that there are a finite set of alternatives
$\mathbb{X}=\{x_{1}, x_{2}, \cdots x_{n}\}$ and a criterion $C$ in a decision-making problem. The binary relation of alternatives with respect to $C$ can be quantified by defining a mapping as follows:
\begin{equation}
P_{C}: \mathbb{X}\times \mathbb{X}\mapsto \mathbb{R}^{+},
\end{equation}
where $\mathbb{R}^{+}$ stands for the set of positive real numbers. Then the fundamental scale in relative measurement satisfies the following rules \citep{Saaty86}:
\begin{description}
\item[(I)] $P_{C}(x_{i},x_{j})=a_{ij}\in \mathbb{R}^{+};$
\item[(II)] $a_{ij}>1$ if $x_{i}$ is strictly preferred to $x_{j}$ under the criterion $C,$ or $x_{i}\succ_{C} x_{j};$
\item[(III)] $a_{ij}=1$ if $x_{i}$ is equivalent to $x_{j}$ under the criterion $C,$ or $x_{i}\sim_{C} x_{j};$
\end{description}
with $\forall i,j\in \mathbb{I}.$ In particular, one has $a_{ii}=1$ since $x_{i}$ is equivalent to $x_{i}.$ Some discussions about the existence of the ratio scales have been widely made \citep{Saaty1977,Dyer1990a,Bernasconi2010}. Here we follow the observations in \cite{Bernasconi2010} that the ratio scales could be derived according to the modern theory of subjective measurement.

\subsection{The initial axiomatic foundation}

In the following, let us recall the four axioms defined by \cite{Saaty86} in terms of the above primitive notions. The basic one is the reciprocal property of pairwise comparisons:
\begin{repeattheorem}[Axiom 1.]
(Reciprocal property) The preference intensities between the alternatives $x_{i}$ and $x_{j}$ satisfy the following relation:
\begin{equation}
a_{ij}=1/a_{ji},\hspace{0.5cm}i,j\in \mathbb{I}. \label{eq2}
\end{equation}
\end{repeattheorem}
When explaining the reciprocal property (\ref{eq2}), \cite{Saaty86} stated that
"{\it if one stone is judged to be five times heavier than another, then the other is automatically one fifth as heavy as the first because it participated in making the first judgment.}"
It is seen that the reciprocal property is mainly based on the mathematical intuition. Moreover, from the theory of subjective measurement in mathematical psychology \citep{Bernasconi2010}, the reciprocal property is not necessarily satisfied.
The main reason is based on the fact that when separately evaluating the preference intensities $a_{ij}$ and $a_{ji},$ some uncertainty could be existing from the viewpoint of human psychology. In addition, to cope with the uncertainty experienced by the decision maker, the interval judgements have been proposed by \cite{Saaty87} to evaluate pairwise comparisons.
In this study, it will be proved that the pairwise comparisons without reciprocal property are equivalent to the interval judgements in a sense.
Therefore, when the reciprocal property is considered to be not necessary, the axiomatic foundation of the AHP model should be modified.

On the other hand, within the framework of the AHP model, a complex decision making problem is decomposed as a hierarchy structure with criteria, subcriteria and alternatives. The hierarchic axioms form the bases of the hierarchy structure. First, the concept of a partially ordered set should be mentioned:
\begin{definition}
A set $\mathbb{S}$ with a binary relation $"\preceq"$ is partially ordered, when the following conditions are satisfied:
\begin{itemize}
  \item Reflexive: $x\preceq x$ for $\forall x\in \mathbb{S};$
  \item Transitive: If $x\preceq y$ and $y\preceq z,$ then $x\preceq z$ for $\forall x,y,x\in \mathbb{S};$
  \item Antisymmetric: If $x\preceq y$ and $y\preceq x,$ then $x\sim y$ for $\forall x,y\in \mathbb{S}.$
\end{itemize}
\end{definition}
Then the boundedness, the supremum and the infimum of a partially ordered set can be further defined. Following the concept of the partially ordered set,
the definition of a hierarchy is given as follows \citep{Saaty86}:
\begin{definition}
A hierarchy $\mathbb{H}$ satisfies the following conditions:
\begin{itemize}
  \item $\mathbb{H}$ is a finite partially ordered set with the largest element $b.$
  \item $\mathbb{H}$ can be partitioned as $h$ subsets called levels $\{\mathbb{L}_{k}, k=1,2,\cdots, h\}$ with $\mathbb{L}_{1}=\{b\}.$
  \item If $x\in \mathbb{L}_{k},$ one has $\mathbb{L}_{x}^{-}=\{y|y\prec x\}\subseteq \mathbb{L}_{k+1}$ $(k=1,2,\cdots,h-1)$ and  $\mathbb{L}_{x}^{+}=\{y|x\prec y \}\subseteq \mathbb{L}_{k-1}$ $(k=2,3,\cdots,h),$
  where the symbol $"\prec"$ stands for "less preferred than."
\end{itemize}
\end{definition}
In connect with the AHP model, the second axiom is given as the following form:
\begin{repeattheorem}[Axiom 2.]
($\rho$-homogeneity) Assume that  $\rho\geq1$ is a positive real number and $\mathbb{H}$ is a hierarchy. $\mathbb{L}_{x}^{-}\subseteq \mathbb{L}_{k+1}$ is of $\rho$-homogeneity with respect to $x\in\mathbb{L}_{k}\subseteq\mathbb{H}$ for $k=1,2,\cdots,h-1$ under the following condition:
\begin{equation}
1/\rho\leq P_{C}(y_{1},y_{2})\leq\rho,\hspace{0.5cm}\forall y_{1}, y_{2}\in \mathbb{L}_{x}^{-}. \label{eq3}
\end{equation}
\end{repeattheorem}
The $\rho$-homogeneity characterizes the comparability of similar things belonging to the same level. When the reciprocal property is not necessary, the condition (\ref{eq3}) should be changed correspondingly.

Furthermore, it is seen that the dependence of different levels should be considered by extending the notions of the fundamental scale.
The concepts of outer dependent and inner dependent are introduced as follows:
\begin{definition}
A set $\mathbb{A}$ is outer dependent on the set $\mathbb{C}$ if one can define a fundamental scale on $\mathbb{A}$ with respect to each $C\in\mathbb{C}.$
\end{definition}
\begin{definition}
Assume that $\mathbb{A}$ is outer dependent on the set $\mathbb{C}.$ The elements in $\mathbb{A}$ are inner dependent with respect to  $C\in\mathbb{C}$
if for some $A\in\mathbb{A},$ $\mathbb{A}$ is outer dependent on $A.$
\end{definition}

The relation between the levels $\mathbb{L}_{k}$ and $\mathbb{L}_{k+1}$ should satisfy the following axiom:
\begin{repeattheorem}[Axiom 3.]
(Dependence) Let $\mathbb{L}_{k}$ ($k=1,2,\cdots,h$) be the levels of a hierarchy $\mathbb{H}.$ Then $\mathbb{L}_{k}$ and $\mathbb{L}_{k+1}$ ($k=1,2,\cdots,h-1$) have the following dependence relations:
\begin{itemize}
  \item $\mathbb{L}_{k+1}$ is outer dependent on $\mathbb{L}_{k};$
  \item $\mathbb{L}_{k+1}$ is not inner dependent with respect to all $x\in\mathbb{L}_{k};$
  \item $\mathbb{L}_{k}$ is not outer dependent on $\mathbb{L}_{k+1}.$
\end{itemize}
\end{repeattheorem}
Axiom 3 is not related to the reciprocal property; and it holds when pairwise comparisons are not reciprocal.

At the end, it is considered that the decision maker usually has an expectation about the outcome of a decision-making problem. The constructed hierarchy should be compatible with the expectation. Hence, the fourth axiom is expressed as follows:
\begin{repeattheorem}[Axiom 4.]
(Expectation) The constructed hierarchy includes all criteria and alternatives; and the derived priorities are compatible with the expectations represented in the hierarchical structure.
\end{repeattheorem}

The above four axioms provide the theoretical basis of the typical AHP choice model \citep{Saaty86}.
In what follows, we mainly modify the first axiom about the reciprocal property of pairwise comparisons to establish a new axiomatic foundation of the AHP model under uncertainty.

\subsection{A modified axiomatic foundation}

As shown in \cite{Harker1987}, Axiom 1 is based on the mathematical intuition where the relations of $x_{i}=a_{ij}x_{j}$ and $x_{j}=x_{i}/a_{ij}$ should be simultaneously satisfied for $a_{ij}>0.$
However, the reciprocal property (\ref{eq2}) could not always hold for a practical case, which can be explained from the two views of point. One is
based on the paired technique of comparing alternatives. When the decision maker evaluates the preference intensities of the alternatives $x_{i}$ over $x_{j}$ and $x_{j}$ over $x_{i},$
the uncertainty could be experienced \citep{Saaty87}. In fact, the reciprocal property (\ref{eq2}) reflects the strict logical relation between $a_{ij}$ and $a_{ji},$ which is incompatible with the uncertainty being experienced by the decision maker. The other comes with the subjective measurement of the ratio scales \citep{Bernasconi2010}. Some cognitive distortions of the decision maker could affect the evaluation of her/his ratio judgements.
Therefore, the reciprocal property (\ref{eq2}) should be relaxed to naturally capture the uncertainty exhibited in paired comparisons.

When softening the equality (\ref{eq2}), there are two situations to be considered. One is $0<a_{ij}a_{ji}\leq1$ and the other is $a_{ij}a_{ji}\geq1.$
For the two considered situations, we can introduce a factor $\theta_{ij}$ such that $\theta_{ij}=a_{ij}a_{ji}.$ Hence the concept of reciprocal symmetry breaking is proposed in the following definition:
\begin{definition}
Suppose that there are a finite set of alternatives $\mathbb{X}=\{x_{1}, x_{2}, \cdots, x_{n}\}.$ The symbols $a_{ij}$ and $a_{ji}$ stand for the preference intensities of $x_{i}$ over $x_{j}$ and $x_{j}$ over $x_{i},$ respectively. If there exist a pair of preference intensities $a_{ij}$ and $a_{ji}$ such that
\begin{equation}
\theta_{ij}=a_{ij}a_{ji}\neq1,
\end{equation}
the pairwise comparisons on $\mathbb{X}$ are of reciprocal symmetry breaking.
\end{definition}

In general, we can consider that the reciprocal symmetry breaking reflects some uncertainty in pairwisely comparing alternatives. In addition, it is interesting to recall the interval-valued comparison matrix provided by \cite{Saaty87}:

\begin{definition}\label{def6}
An interval-valued comparison matrix is represented as:
\begin{equation}
\tilde{A}=(\tilde{a}_{ij})_{n\times n}=\left(
\begin{array}{cccc}
\left[{1,1}\right]&{\left[{a^-_{12},a^+_{12}}\right]}&\cdots&{\left[{a^-_{1n},a^+_{1n}}\right]}\\
\left[{a^-_{21},a^+_{21}}\right]&{\left[{1,1}\right]}&\cdots&\left[{a^-_{2n},a^+_{2n}}\right]\\
\cdots&\cdots&\cdots&\cdots\\
\left[{a^-_{n1}, a^+_{n1}}\right]&\left[{a^-_{n2}, a^+_{n2}}\right]&\cdots&{\left[{1,1}\right]}
\end{array}\right).
\end{equation}
Hereafter $\tilde{a}_{ij}=\left[{a^-_{ij}, a^+_{ij}}\right]$ means that the preference intensity of $x_{i}$ over $x_{j}$ is located between $a^-_{ij}$ and $ a^+_{ij}.$ The values satisfy
${a^{-}_{ij}\cdot a^{+}_{ji}}={a^{+}_{ij}\cdot a^{-}_{ji}}=1$ and $0<a^{-}_{ij}\leq a^{+}_{ij}$ for $i, j\in \mathbb{I}.$
\end{definition}

\begin{figure}[t]
\begin{center}
\includegraphics[height=1.6in]{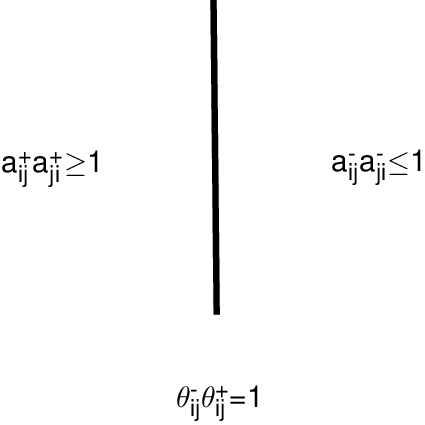}
\caption{The mirror mechanism between the cases of $0<a_{ij}^{-}a_{ji}^{-}\leq 1$ and $a_{ij}^{+}a_{ji}^{+}\geq1.$ } \label{figure2}
\end{center}
\end{figure}

Definition \ref{def6} shows the following relations:
\begin{equation}
a_{ij}^{-}a_{ji}^{-}\leq1,\quad a_{ij}^{+}a_{ji}^{+}\geq1,\quad i,j\in \mathbb{I}.
\end{equation}
Moreover, letting
\begin{equation}
\theta_{ij}^{-}=a_{ij}^{-}a_{ji}^{-}, \quad \theta_{ij}^{+}=a_{ij}^{+}a_{ji}^{+},
\end{equation}
we have
\begin{equation}
\theta_{ij}^{-}\theta_{ij}^{+}=1.\label{eq8}
\end{equation}
As shown in Figure \ref{figure2}, the mirror mechanism has been built for the transformation between the cases of $0<a_{ij}^{-}a_{ji}^{-}\leq 1$ and $a_{ij}^{+}a_{ji}^{+}\geq1.$ In other words, when we give $a_{ij}^{+}$ and $a_{ji}^{+}$ with $a_{ij}^{+}a_{ji}^{+}\geq1$ by comparing $x_{i}$ and $x_{j},$ the comparison ratios $a_{ij}^{-}$ and $a_{ji}^{-}$ with $0<a_{ij}^{-}a_{ji}^{-}\leq1$ have been determined by using the relation ${a^{-}_{ij}\cdot a^{+}_{ji}}={a^{+}_{ij}\cdot a^{-}_{ji}}=1.$
This means that it is sufficient to consider one of the two cases with $0<\theta_{ij}^{-}\leq 1$ and $\theta_{ij}^{+}\geq1,$ where the
reciprocal symmetry (\ref{eq8}) is satisfied. In order to form an axiomatic foundation of the AHP model under some uncertainty, Axiom 1 is modified as follows:
\begin{repeattheorem}[Axiom 1$^\prime$.]
(Reciprocal symmetry breaking) The preference intensities of the alternatives $x_{i}$ over $x_{j}$ and $x_{j}$ over $x_{i}$ satisfy the following relation:
\begin{equation}
0<a_{ij}a_{ji}\leq1,\hspace{0.5cm}i,j\in \mathbb{I}. \label{eq9}
\end{equation}
\end{repeattheorem}
Clearly, Axiom 1 is the particular case of Axiom 1$^\prime$ with reciprocal property. In a similar manner, Axiom 2 should be further adjusted and we give the following axiom:
\begin{repeattheorem}[Axiom 2$^\prime$.]
($\rho$-homogeneity) Let  $\rho\geq1$ and $\mathbb{H}$ be a positive real number and a hierarchy, respectively. $\mathbb{L}_{x}^{-}\subseteq \mathbb{L}_{k+1}$ is of $\rho$-homogeneity with respect to $x\in\mathbb{L}_{k}\subseteq\mathbb{H}$ for $k=1,2,\cdots,h-1$ and the following conditions:
\begin{equation}
1/\rho\leq P_{C}(y_{1},y_{2})\leq\rho,\quad 1/\rho\leq P_{C}(y_{2},y_{1})\leq\rho\hspace{0.5cm}\forall y_{1}, y_{2}\in \mathbb{L}_{x}^{-}. \label{eq10}
\end{equation}
\end{repeattheorem}
As compared to Axiom 2, the pairwise comparisons $P_{C}(y_{1},y_{2})$ and $P_{C}(y_{2},y_{1})$ are all considered in Axiom 2$^\prime$.
Furthermore, since Axiom 3 is only related to the fundamental scale satisfying the three rules (I)-(III), it does not need to be modified. Additionally, Axiom 4 holds, even when
the irrational behavior of the decision maker emerges \citep{Saaty86}. In a word, Axioms 1$^\prime$, 2$^\prime$, 3 and 4 form the modified axiomatic foundation of the AHP model under some uncertainty.

\section{Results from the modified axioms}

It is noted from the results in \cite{Saaty86} that the reciprocal property is the necessary condition of a consistent binary relation.
When the reciprocal property is broken, the pairwise comparisons must be inconsistent.
This implies that the inconsistency is the natural property of pairwise comparisons according to Axiom 1$^\prime$.
The above observation is in agreement with the practical situation, since one always provides
inconsistent judgements when pairwisely comparing alternatives \citep{Saaty80}.
In the following, we derive some interesting results from the modified axiomatic foundation.

\subsection{Properties of reciprocal symmetry breaking}

It is convenient to propose the concept of pairwise comparison matrices with reciprocal symmetry breaking.

\begin{definition}\label{def7}
A pairwise comparison matrix $A=(a_{ij})_{n\times n}$ is of reciprocal symmetry breaking, if the derived matrix
\begin{equation}
\Theta=(\theta_{ij})_{n\times n}\neq E,\label{eq11}
\end{equation}
where
$\theta_{ij}=a_{ij}a_{ji}$ ($i,j\in \mathbb{I}$)
and $E$ stands for the matrix whose entries are all ones.
\end{definition}

Hereafter, when we say pairwise comparison matrices, it means that the reciprocal property could be breaking. In terms of (\ref{eq11}), it gives $\theta_{ij}=\theta_{ji}$ ($i,j\in \mathbb{I}$), meaning that the constructed matrix $\Theta$ is symmetrical.
When $\Theta=E,$ the matrix $A=(a_{ij})_{n\times n}$ is with reciprocal property. This implies that the matrix $\Theta$ characterizes the reciprocal property of a
pairwise comparison matrix. For the sake of distinguishing, we give the following definition:
\begin{definition}\label{def8}
If a pairwise comparison matrix $A=(a_{ij})_{n\times n}$ is with reciprocal property, it is called a multiplicative reciprocal matrix.
\end{definition}

Moreover, the degree of reciprocal symmetry breaking of a pairwise comparison matrix $A=(a_{ij})_{n\times n}$ can be quantified by using the matrix $\Theta.$ We define the following index:
\begin{definition}\label{def9}
Suppose that $A=(a_{ij})_{n\times n}$ is a pairwise comparison matrix. The degree of reciprocal symmetry breaking is defined by the following equality:
\begin{equation}
SBD(A)=\frac{2}{n(n-1)}\sum_{i=1}^{n}\sum_{j=i+1}^{n}\theta_{ij},\quad n\geq2,\label{eq12}
\end{equation}
where $\theta_{ij}=a_{ij}a_{ji}$ for $i,j\in \mathbb{I}.$
\end{definition}

In addition, the index of $SBD(A)$ has the property as follows:
\begin{theorem}\label{th1}
The degree of reciprocal symmetry breaking of a pairwise comparison matrix $A=(a_{ij})_{n\times n}$ satisfies the relation of $0<SBD(A)\leq1.$
$A=(a_{ij})_{n\times n}$ is a multiplicative reciprocal matrix if and only if $SBD(A)=1.$
\end{theorem}
\proof{Proof of Theorem \ref{th1}.}
In terms of Axiom 1$^\prime$ and Definition \ref{def9}, it follows $0<\theta_{ij}\leq1$ and $\theta_{ij}=\theta_{ji}$ for $i,j\in \mathbb{I}.$
We further have
\begin{equation}
0<SBD(A)=\frac{2}{n(n-1)}\sum_{i=1}^{n}\sum_{j=i+1}^{n}\theta_{ij}\leq1.
\end{equation}
If $SBD(A)=1,$ then $\theta_{ij}=1$ for $i,j\in \mathbb{I}$ and $j>i.$ In virtue of $\theta_{ij}=\theta_{ji}$ and $\theta_{ii}=1,$ one has $\Theta=(\theta_{ij})_{n\times n}=E,$ meaning that
$A=(a_{ij})_{n\times n}$ is a multiplicative reciprocal matrix (Definition 8). On the other hand, if $A=(a_{ij})_{n\times n}$ is with reciprocal property, then $\theta_{ij}=a_{ij}a_{ji}=1.$ The application of (\ref{eq12}) leads to $SBD(A)=1.$
\endproof

Under Axiom 1$^\prime$, the proposed index $SBD(A)$ reflects the breaking degree of reciprocal property of a pairwise comparison matrix. The more the value of $SBD(A),$ the less the breaking degree of reciprocal property is. It can be further explained as the subjective probability to give a pairwise comparison matrix with reciprocal property from the viewpoint of the philosophical discussion. In other words, when $SBD(A)=1,$ it implies that the subjective probability of giving a pairwise comparison matrix with reciprocal property is 1. With the decreasing of the value of $SBD(A),$ the subjective probability of providing a multiplicative reciprocal matrix is decreasing.
Furthermore, since an interval-valued comparison matrix can be used to capture the uncertainty experienced by the decision maker \citep{Saaty87}, the following result is achieved:
\begin{theorem}\label{th2}
A pairwise comparison matrix $A=(a_{ij})_{n\times n}$ with reciprocal symmetry breaking is equivalent to an interval-valued
comparison matrix $\tilde{A}=(\tilde{a}_{ij})_{n\times n}.$
\end{theorem}
\proof{Proof of Theorem \ref{th2}.} If $A=(a_{ij})_{n\times n}$ is with reciprocal symmetry breaking,  we have
$0<\theta_{ij}=a_{ij}a_{ji}\leq1$ under Axiom 1$^\prime$ for $\forall i,j\in \mathbb{I}$ and $\Theta\neq E$ according to Definition \ref{def7}.
Then it follows
\begin{equation}
a_{ij}\leq \frac{1}{a_{ji}}.
\end{equation}
Letting
$$a_{ij}^{-}=a_{ij},\hspace{0.5cm} a_{ij}^{+}=1/a_{ji},$$
and
$$\tilde{a}_{ij}=[a_{ij}^{-},a_{ij}^{+}],$$
one obtains the interval-valued comparison matrix $\tilde{A}=(\tilde{a}_{ij})_{n\times n}.$

On the contrary, when an interval-valued comparison matrix $\tilde{A}=(\tilde{a}_{ij})_{n\times n}$ with $\tilde{a}_{ij}=[a_{ij}^{-},a_{ij}^{+}]$ is given,
the pairwise comparison matrix $A=(a_{ij})_{n\times n}$ with reciprocal symmetry breaking is derived by assuming $a_{ij}=a_{ij}^{-}$ for $i,j\in \mathbb{I}$ under the consideration of Axiom 1$^\prime$.

\endproof

The equivalence shown in Theorem 2 reveals that the uncertainty exhibited in an interval-valued comparison matrix can be equivalently captured by a pairwise comparison matrix with reciprocal symmetry breaking. In addition, following the idea in Definition \ref{def9}, an uncertainty index of $\tilde{A}=(\tilde{a}_{ij})_{n\times n}$ can be constructed as
\begin{equation}
UI(\tilde{A})=\frac{2}{n(n-1)}\sum_{i=1}^{n}\sum_{j>i}^{n}a_{ij}^{-}a_{ji}^{-}. \label{eq15}
\end{equation}
Then the following result is obtained:
\begin{theorem}\label{th3}
Assume that $\tilde{A}=(\tilde{a}_{ij})_{n\times n}$ is an interval-valued comparison matrix with $\tilde{a}_{ij}=[a_{ij}^{-},a_{ij}^{+}]$ for $i,j\in \mathbb{I}.$
$\tilde{A}=(\tilde{a}_{ij})_{n\times n}$ degenerates to a multiplicative reciprocal matrix if and only if the value of the constructed uncertainty index $UI(\tilde{A})$ is equal to 1.
\end{theorem}
\proof{Proof of Theorem \ref{th3}.}
On the one hand, if $\tilde{A}=(\tilde{a}_{ij})_{n\times n}$ degenerates to a multiplicative reciprocal matrix, then $a_{ij}^{-}=a_{ij}^{+}$ and $a_{ij}^{-}a_{ji}^{-}=1$ for $i,j\in \mathbb{I}.$ Using the formula (\ref{eq15}), one has $UI(\tilde{A})=1.$ On the other hand, if $UI(\tilde{A})=1,$ the result of Theorem \ref{th1} shows that the matrix
$A^{L}=(a_{ij}^{-})_{n\times n}$ is a multiplicative reciprocal matrix. This means that $a_{ij}^{-}a_{ji}^{-}=1$ and $a_{ij}^{-}=1/a_{ji}^{-}=a_{ij}^{+}.$ That is, $\tilde{A}=(\tilde{a}_{ij})_{n\times n}$ degenerates to a multiplicative reciprocal matrix $A^{L}.$

\endproof

It is seen from the above observations that the uncertainty experienced by the decision maker has been naturally incorporated into Axiom 1$^\prime$. Therefore,
the modified axiomatic foundation can be considered as the basis of the AHP model under some uncertainty.

\subsection{Approximate consistency of pairwise comparison matrices}

It is worth noting that the concept of consistent pairwise comparison matrices is important and it is recalled as follows \citep{Saaty80,Saaty86}:
\begin{definition}\label{dj10}
A pairwise comparison matrix $A=(a_{ij})_{n\times n}$ is consistent if
\begin{equation}
a_{ij}a_{jk}=a_{ik}, \quad i,j,k\in \mathbb{I}.
\end{equation}
\end{definition}
In this study, it is noted that a pairwise comparison matrix could be with the breaking of reciprocal property. Hence, the consistency is only a particular property of a pairwise comparison matrix and a softened version of consistency should be further developed. It is noted that the concept of approximate consistency has been proposed to characterize the consistency property of interval-valued comparison matrices \citep{LiuF2017}. Following the idea in \cite{LiuF2017}, the concept of approximate consistency of a pairwise comparison matrix is proposed:
\begin{definition}\label{dj11}
A pairwise comparison matrix $A=(a_{ij})_{n\times n}$ is approximately consistent, if the same ranking of alternatives can be obtained by using each row and column vectors in $A=(a_{ij})_{n\times n}.$
\end{definition}

In Definition \ref{dj11}, the restrictive quantitative relation among paired comparisons in Definition \ref{dj10} has been neglected and the ranking of alternatives is directly used.
For example, let us consider a pairwise comparison matrix as follows \citep{Saaty2013}:
$$
A_{1}=\left(
\begin{array}{c|ccccc}
C&x_{1}&x_{2}&x_{3}&x_{4}&x_{5}\\
\hline
x_{1}&1&9&2&3&5\\
x_{2}&1/9&1&1/5&1/3&1/2\\
x_{3}&1/2&5&1&3/2&3\\
x_{4}&1/3&3&2/3&1&3/2\\
x_{5}&1/5&2&1/3&2/3&1\\
\end{array}
\right).
$$
It is easy to determine the same ranking of alternatives as $x_{1}\succ x_{3}\succ x_{4}\succ x_{5}\succ x_{2}$ by using each row and column vectors in $A_{1},$
meaning that $A_{1}$ is of approximate consistency according to Definition \ref{dj11}. In addition, one can see that $A_{1}$ is with reciprocal property. For the purpose of agreeing with Axiom 1$^\prime$, we further modify the matrix $A_{1}$ as:
$$
A_{1}^{m}=\left(
\begin{array}{c|ccccc}
C&x_{1}&x_{2}&x_{3}&x_{4}&x_{5}\\
\hline
x_{1}&1&\mathbf{8}&2&3&5\\
x_{2}&1/9&1&1/5&1/3&1/2\\
x_{3}&1/2&5&1&3/2&3\\
x_{4}&1/3&3&2/3&1&3/2\\
x_{5}&1/5&2&1/3&2/3&1\\
\end{array}
\right),
$$
where $a_{12}a_{21}=8/9<1.$ The determined ranking of alternatives by using each row and column in $A_{1}^{m}$ still is $x_{1}\succ x_{3}\succ x_{4}\succ x_{5}\succ x_{2},$ meaning that
$A_{1}^{m}$ is also a pairwise comparison matrix with approximate consistency. Hence, due to the breaking of reciprocal property in $A_{1}^{m},$ the concept of approximate consistency is a softened version of consistency, and it is compatible with  Axiom 1$^\prime$. According to the standard of choosing the best alternative, the two matrices $A_{1}$ and $A_{1}^{m}$ are equivalent.
Furthermore, we have the following property:
\begin{theorem}\label{th4}
Let $A=(a_{ij})_{n\times n}$ be a pairwise comparison matrix with approximate consistency. There is a permutation $\sigma: \mathbb{I}\mapsto \mathbb{I}$ such that
$A_{\sigma}=(a_{\sigma(i)\sigma(j)})_{n\times n}$ has the following properties:
\begin{enumerate}
  \item[(1)] $A_{\sigma}$ is a pairwise comparison matrix with approximate consistency;
  \item[(2)] The elements in each row of $A_{\sigma}$ are ranked by the ascending order;
  \item[(3)] The elements in each column of $A_{\sigma}$ are ranked by the descending order.
\end{enumerate}
\end{theorem}
\proof{Proof of Theorem \ref{th4}.}
(1) Suppose that $A=(a_{ij})_{n\times n}$ is with approximate consistency. For any permutation $\sigma,$ there is a unique number $i_{k}\in \mathbb{I}$ such that $\sigma(i)=i_{k}$ for $\forall i\in \mathbb{I}.$ When $A_{\sigma}=(a_{\sigma(i)\sigma(j)})_{n\times n}$ is obtained by applying a permutation $\sigma$ to $A=(a_{ij})_{n\times n},$ the entries in the $i$th row or the $j$th column belonging to $A=(a_{ij})_{n\times n}$ are changed to the $\sigma(i)$th row or the $\sigma(j)$ column in $A_{\sigma}=(a_{\sigma(i)\sigma(j)})_{n\times n}.$ This means that the ranking of alternatives determined by a row or column of $A=(a_{ij})_{n\times n}$ is not changed with respect to the permutation $\sigma.$ Therefore, $A_{\sigma}=(a_{\sigma(i)\sigma(j)})_{n\times n}$ is of approximate consistency.

(2) For $A=(a_{ij})_{n\times n}$ with approximate consistency, it is assumed that the determined ranking of alternatives is $x_{\sigma(1)}\succeq x_{\sigma(2)}\succeq\cdots\succeq x_{\sigma(n)},$ where the permutation $\sigma$ is given. Then applying the permutation $\sigma$ to $A=(a_{ij})_{n\times n}$ to give $A_{\sigma}=(a_{\sigma(i)\sigma(j)})_{n\times n},$
we have $a_{\sigma(i)\sigma(1)}\leq a_{\sigma(i)\sigma(2)}\leq\cdots\leq a_{\sigma(i)\sigma(n)}$ for $i\in \mathbb{I}$ when the ranking is obtained by using the row vectors. This means that the elements in each row of $A_{\sigma}$ are ranked by the ascending order.

(3) Based on the findings in (2), when the ranking of $x_{\sigma(1)}\succeq x_{\sigma(2)}\succeq\cdots\succeq x_{\sigma(n)}$ is determined by using the column vectors, it follows
 $a_{\sigma(1)\sigma(j)}\geq a_{\sigma(2)\sigma(j)}\geq\cdots\geq a_{\sigma(n)\sigma(j)}$ for $j\in \mathbb{I}.$ That is, the elements in each column of $A_{\sigma}$ are ranked by the descending order.

\endproof

Theorem \ref{th4} can be verified by adjusting $A_{1}^{m}$ as the following form:
$$
A_{1}^{\sigma}=\left(
\begin{array}{c|ccccc}
C&x_{1}&x_{3}&x_{4}&x_{5}&x_{2}\\
\hline
x_{1}&1&2&3&5&\mathbf{8}\\
x_{3}&1/2&1&3/2&3&5\\
x_{4}&1/3&2/3&1&3/2&3\\
x_{5}&1/5&1/3&2/3&1&2\\
x_{2}&1/9&1/5&1/3&1/2&1\\
\end{array}
\right).
$$
It is easy to see that the matrix $A_{1}^{\sigma}$ is in agreement with Theorem \ref{th4}. In addition, we have the following observation:
\begin{theorem}\label{th5}
If a pairwise comparison matrix $A=(a_{ij})_{n\times n}$ is consistent (Definition \ref{dj10}), it is of approximate consistency (Definition \ref{dj11}).
\end{theorem}
\proof{Proof of Theorem \ref{th5}.}
Suppose that $A=(a_{ij})_{n\times n}$ is a consistent matrix. Then one has $rank(A)=1$ and all rows of $A=(a_{ij})_{n\times n}$ are identical except for a constant factor \citep{Saaty86}.
That is, when a row of $A=(a_{ij})_{n\times n}$ is used to obtain the ranking of alternatives, the other rows can be used to determine the same ranking of alternatives.
Similarly, when considering the columns of $A=(a_{ij})_{n\times n},$ the same result can be found. Furthermore, due to the reciprocal property of a consistent matrix, the rankings of alternatives using the rows and columns are identical. Therefore, a consistent pairwise comparison matrix is with approximate consistency.

\endproof

The above observations show that a consistent pairwise comparison matrix is the ideal case under the typical and modified axiomatic foundations of the AHP model. A pairwise comparison matrix with approximate consistency can be considered as the ideal case of the modified axiomatic foundation of the AHP model under some uncertainty.

\subsection{Priorities of alternatives}

Another important problem is how to elicit the priorities of alternatives from a pairwise comparison matrix $A=(a_{ij})_{n\times n}$ in the AHP model.
Following the idea in \cite{Saaty80}, let us assume that there is a mapping given as follows:
\begin{equation}
\psi: \mathbb{R}_{M(n)}\mapsto [0,1]^{n},
\end{equation}
where $\mathbb{R}_{M(n)}$ denotes the set of pairwise comparison matrices $A=(a_{ij})_{n\times n}$ and $[0, 1]^{n}$ stands for the $n$-fold Cartesian product of $[0, 1].$
For convenience, the priority vector of alternatives is expressed as $\omega=(\omega_{1}, \omega_{2}, \cdots, \omega_{n})\in [0, 1]^{n}$ with $\sum_{i=1}^{n}\omega_{i}=1.$
The methods should be studied for deriving the priority vector of alternatives from a pairwise comparison matrix. As shown in \cite{Saaty80,Saaty86}, the eigenvalue method has been carefully discussed, especially for consistent pairwise comparison matrices. Here we mainly focus on the method of deriving the priority vector from a pairwise comparison matrix with approximate consistency.

Suppose that the set of pairwise comparison matrices with approximate consistency is written as $\mathbb{R}_{AC(n)}.$ The following result is given:
\begin{theorem}\label{th6}
Let $A=(a_{ij})_{n\times n}\in \mathbb{R}_{AC(n)}$ be obtained by pairwise comparing alternatives in $\mathbb{X}=\{x_{1}, x_{2}, \cdots, x_{n}\}.$ There exists a mapping:
$$
\psi: A\in \mathbb{R}_{AC(n)}\mapsto [0,1]^{n}\ni \omega=(\omega_{1}, \omega_{2}, \cdots, \omega_{n}),
$$
such that
\begin{enumerate}
  \item[(I)] $a_{ij}=\varepsilon_{ij}\frac{\omega_{i}}{\omega_{j}}$ with $\varepsilon_{ij}=\varepsilon_{ji}=\sqrt{a_{ij}a_{ji}}\in(0,1]$ for $i,j\in \mathbb{I}.$
  \item[(II)] For any two alternatives $x_{i}$ and $x_{j},$ $x_{i}\succeq x_{j}$ if and only if $\omega_{i}\geq\omega_{j}$ for $i,j\in \mathbb{I}.$
\end{enumerate}
\end{theorem}
\proof{Proof of Theorem \ref{th6}.}
For $A=(a_{ij})_{n\times n}\in \mathbb{R}_{AC(n)},$ there is a permutation $\sigma$ such that the derived matrix $A_{\sigma}=(a_{\sigma(i)\sigma(j)})_{n\times n}$ satisfies Theorem \ref{th4}.
Without loss of generality, it is assumed that $\sigma=(1,2,\cdots,n)$ and $A=(a_{ij})_{n\times n}$ satisfies Theorem \ref{th4}.
This means that $x_{1}\succeq x_{2} \succeq\cdots\succeq x_{n}$ along with $a_{i1}\leq a_{i2}\leq\cdots\leq a_{in}$ and $a_{1j}\geq a_{2j}\geq\cdots\geq a_{nj}$ for $i,j\in I.$
Therefore, there exists a weight $\omega_{i}$ for the alternative $x_{i}$ with $i\in \mathbb{I}$ such that $\omega_{1}\geq\omega_{2}\geq\cdots\geq\omega_{n}.$ Moreover, a pairwise comparison matrix can be constructed as follows:
\begin{equation}
\Omega=\left(
\begin{array}{c|cccc}
C&x_{1}&x_{2}&\cdots&x_{n}\\
\hline
x_{1}&\frac{\omega_{1}}{\omega_{1}}&\varepsilon_{12}\frac{\omega_{1}}{\omega_{2}}&\cdots&\varepsilon_{1n}\frac{\omega_{1}}{\omega_{n}}\\
x_{2}&\varepsilon_{21}\frac{\omega_{2}}{\omega_{1}}&\frac{\omega_{2}}{\omega_{2}}&\cdots&\varepsilon_{2n}\frac{\omega_{2}}{\omega_{n}}\\
\vdots&\vdots&\vdots&\vdots&\vdots\\
x_{n}&\varepsilon_{n1}\frac{\omega_{n}}{\omega_{1}}&\varepsilon_{n2}\frac{\omega_{n}}{\omega_{2}}&\cdots&\frac{\omega_{n}}{\omega_{n}}\\
\end{array}
\right).\label{eq18}
\end{equation}
Letting
$$
a_{ij}=\varepsilon_{ij}\frac{\omega_{i}}{\omega_{j}},\quad \varepsilon_{ij}=\varepsilon_{ji},
$$
we have
$$
A=\Omega,\quad a_{ij}a_{ji}=\varepsilon_{ij}^{2}.
$$
In terms of Axiom 1$^\prime$, it gives $\varepsilon_{ij}\in(0,1],$
then the results (I) and (II) follow.

\endproof

In what follows, we prove that the eigenvalue method still is feasible to derive the priority vector from a pairwise comparison matrix with approximate consistency.
It is convenient to recall a lemma as follows:

\begin{lemma} \label{lem1}
Let $A=(a_{ij})_{n\times n}$ be a pairwise comparison matrix and $\sigma$ be a permutation of $(1,2,\cdots,n).$ $A_{\sigma}=(a_{\sigma(i)\sigma(j)})_{n\times n}$  is
determined by applying $\sigma$ to $A=(a_{ij})_{n\times n}.$ Then the eigenvalues of $A=(a_{ij})_{n\times n}$ and $A_{\sigma}=(a_{\sigma(i)\sigma(j)})_{n\times n}$ are identical. The corresponding eigenvectors of $A_{\sigma}=(a_{\sigma(i)\sigma(j)})_{n\times n}$ are obtained by applying $\sigma$ to those of $A=(a_{ij})_{n\times n}.$
\end{lemma}

The proof of Lemma \ref{lem1} can be completed by using the matrix theory \citep{HornRA1985} and the known finding \citep{LiuF2020}. The detail procedure has been neglected here.
Furthermore, we have the following result:

\begin{theorem}\label{th7}
Let $A=(a_{ij})_{n\times n}\in \mathbb{R}_{AC(n)}$ be expressed as (\ref{eq18}) and $A^{\prime}$ be a consistent pairwise comparison matrix.
The principal right eigenvector of $A^{\prime}$ is written as $\omega^{\prime}=(\omega_{1}, \omega_{2}, \cdots, \omega_{n})^{T},$ where $T$ denotes the transposition.
The principal right eigenvector of $A$ corresponding to the principal eigenvalue $\lambda_{max}$ is calculated as $\omega_{a}=(\omega_{a1}, \omega_{a2}, \cdots, \omega_{an})^{T}.$ A permutation $\sigma$ is applied to $\omega^{\prime}$ and $\omega_{a}$ to get $\omega^{\prime}_{\sigma}=(\omega_{\sigma(1)}, \omega_{\sigma(2)}, \cdots, \omega_{\sigma(n)})^{T}$ and $\omega_{a\sigma}=(\omega_{a\sigma(1)}, \omega_{a\sigma(2)}, \cdots, \omega_{a\sigma(n)})^{T},$ respectively. Then there is a permutation $\sigma$ such that
$\omega_{a\sigma(1)}\geq\omega_{a\sigma(2)}\geq\cdots\geq\omega_{a\sigma(n)}$ and $\omega_{\sigma(1)}\geq\omega_{\sigma(2)}\geq\cdots\geq\omega_{\sigma(n)}$ are satisfied simultaneously.
\end{theorem}
\proof{Proof of Theorem \ref{th7}.}
 Since $A=(a_{ij})_{n\times n}\in \mathbb{R}_{AC(n)},$ there is a permutation $\sigma$ such that $A_{\sigma}=(a_{\sigma(i)\sigma(j)})_{n\times n}$ with the following relations:
 \begin{equation}
0<a_{\sigma(i)\sigma(j)}\leq a_{\sigma(i)\sigma(j+1)},\quad a_{\sigma(i)\sigma(j)}\geq a_{\sigma(i+1)\sigma(j)}>0,\quad  i,j=1,2,\cdots,n-1,\label{eq19}
 \end{equation}
 where Theorem \ref{th4} has been used. Making use of Lemma \ref{lem1}, we have
\begin{equation}
A_{\sigma}\omega_{a\sigma}=\lambda_{max}\omega_{a\sigma}.
\end{equation}
This means that
\begin{equation}
\sum_{j=1}^{n}a_{\sigma(i)\sigma(j)}\omega_{a\sigma(j)}=\lambda_{max}\omega_{a\sigma(i)},\quad i\in \mathbb{I}.
\end{equation}
In virtue of (\ref{eq19}), it follows $\omega_{a\sigma(1)}\geq\omega_{a\sigma(2)}\geq\cdots\geq\omega_{a\sigma(n)}.$ Moreover, by rewriting $A=(a_{ij})_{n\times n}$ as the matrix in (\ref{eq18}), the application of Theorem \ref{th6} yields $\omega_{\sigma(1)}\geq\omega_{\sigma(2)}\geq\cdots\geq\omega_{\sigma(n)}.$

\endproof

As shown in Theorem \ref{th7}, the eigenvalue method can be used as the method of eliciting the properties of alternatives from a pairwise comparison matrix with approximate consistency. For example, the pairwise comparison matrix $A_{1}^{\sigma}$ with approximate consistency is utilized for numerical computations. The largest eigenvalue and the corresponding eigenvector can be determined as $4.9824$ and
$(0.4565, 0.2476, 0.1523, 0.0940, 0.0496),$ respectively. This means that the ranking of alternatives is $x_{1}\succ x_{3}\succ x_{4}\succ x_{5}\succ x_{2},$ which is in agreement with the existing result. In addition, it is noted that the largest eigenvalue $4.9824$ is less than the order $5$ of the matrix. The observation is attributed to the breaking of reciprocal symmetry and different from the result of \cite{Saaty86}.

\section{Ranking reversal phenomenon}

The phenomenon of ranking reversal has been discussed widely within the framework of the AHP model. However, in the open literature, numerical examples are always offered to illustrate the ranking reversal phenomenon and the theoretical investigation is little made except for the finding in \cite{Saaty1984b}. Here we attempt to give a theoretical analysis and to reach a general result according to the modified axiomatic foundation.

\subsection{Single criterion case}

When a single criterion is considered to compare alternatives in pairs, the rank preservation has been studied in \cite{Saaty1984b}.
It can be resulted that when the pairwise comparison matrix is consistent, the ranking of alternatives is not changed by the action of adding or deleting an alternative \citep{Saaty86}.
In other words, the ranking reversal phenomenon should not appear when the judgements of the decision maker are consistent.
Now we consider the case that the pairwise comparison matrix is of approximate consistency in terms of Definition \ref{dj11}.
\begin{theorem}\label{th8}
Let $A=(a_{ij})_{n\times n}\in \mathbb{R}_{AC(n)}.$ When deleting an alternative, the obtained pairwise comparison matrix is written as $A_{d}=(a_{ij}^{d})_{(n-1)\times (n-1)}.$ We have
$A_{d}\in \mathbb{R}_{AC(n-1)}$ and the ranking of the remaining alternatives holds the same as that determined by $A.$
\end{theorem}
\proof{Proof of Theorem \ref{th8}.}
Since $A=(a_{ij})_{n\times n}$ is a pairwise comparison matrix with approximate consistency, there is a permutation $\sigma$ such that
\begin{equation}
0<a_{\sigma(i)\sigma(j)}\leq a_{\sigma(i)\sigma(j+1)},\quad a_{\sigma(i)\sigma(j)}\geq a_{\sigma(i+1)\sigma(j)}>0,\quad  i,j=1,2,\cdots,n-1.\label{eq22}
 \end{equation}
When deleting any an alternative, the corresponding comparison ratios are deleted. The remaining entries are used to construct $A_{d}$ and their rankings are still kept as (\ref{eq22}), meaning $A_{d}\in \mathbb{R}_{AC(n-1)}.$
Based on Theorems \ref{th6} and \ref{th7}, the ranking of the remaining alternatives is not changed.

\endproof
\begin{theorem}\label{th9}
Let $A=(a_{ij})_{n\times n}\in \mathbb{R}_{AC(n)}.$ A new pairwise comparison matrix is determined as $A_{a}=(a_{ij}^{a})_{(n+1)\times (n+1)}$ when adding an alternative. If
$A_{a}\in \mathbb{R}_{AC(n+1)},$ the ranking of the old alternatives is the same as that obtained by $A.$
\end{theorem}
\proof{Proof of Theorem \ref{th9}.}
For $A=(a_{ij})_{n\times n}\in \mathbb{R}_{AC(n)},$ there exists a permutation $\sigma$ such that
\begin{equation}
x_{\sigma(1)}\succeq x_{\sigma(2)}\succeq \cdots \succeq x_{\sigma(n)},\label{eq23}
\end{equation}
with
\begin{equation}
0<a_{\sigma(i)\sigma(j)}\leq a_{\sigma(i)\sigma(j+1)},\quad a_{\sigma(i)\sigma(j)}\geq a_{\sigma(i+1)\sigma(j)}>0,\quad  i,j=1,2,\cdots,n-1.
 \end{equation}
When adding an alternative $x_{n+1}$ and $A_{a}=(a_{ij}^{a})_{(n+1)\times (n+1)}\in \mathbb{R}_{AC(n+1)},$ there is a permutation $\sigma^\prime$ such that
\begin{equation}
x_{\sigma^\prime(1)}\succeq x_{\sigma^\prime(2)}\succeq \cdots \succeq x_{\sigma^\prime(n+1)}.\label{eq25}
\end{equation}
Obviously, the ranking in (\ref{eq23}) is not changed in (\ref{eq25}).

\endproof

It is convenient to reconsider the matrix $A_{1}^{\sigma}.$ For example, deleting the alternative $x_{5},$ it gives
$$
A_{1}^{d}=\left(
\begin{array}{c|cccc}
C&x_{1}&x_{3}&x_{4}&x_{2}\\
\hline
x_{1}&1&2&3&\mathbf{8}\\
x_{3}&1/2&1&3/2&5\\
x_{4}&1/3&2/3&1&3\\
x_{2}&1/9&1/5&1/3&1\\
\end{array}
\right),
$$
where the elements corresponding to $x_{5}$ in $A_{1}^{\sigma}$ have been deleted. The ranking of alternatives is $x_{1}\succ x_{3} \succ x_{4} \succ x_{2}$ by using $A_{1}^{d},$ which is identical with that given by $A_{1}^{\sigma}.$ When adding an alternative $x_{6}$ to give the matrix with approximate consistency as:
$$
A_{1}^{a}=\left(
\begin{array}{c|cccccc}
C&x_{1}&x_{3}&x_{4}&x_{5}&x_{6}&x_{2}\\
\hline
x_{1}&1&2&3&5&6&\mathbf{8}\\
x_{3}&1/2&1&3/2&3&4&5\\
x_{4}&1/3&2/3&1&3/2&2&3\\
x_{5}&1/5&1/3&2/3&1&3/2&2\\
x_{6}&1/6&1/4&1/2&2/3&1&\mathbf{3/2}\\
x_{2}&1/9&1/5&1/3&1/2&4/5&1\\
\end{array}
\right),
$$
it is easy to obtain the ranking of alternatives as $x_{1}\succ x_{3}\succ x_{4}\succ x_{5}\succ x_{6}\succ x_{2},$ where the ranking of $x_{1}\succ x_{3}\succ x_{4}\succ x_{5}\succ x_{2}$ is not changed. Although pairwise comparison matrices with the breaking of reciprocal property are inconsistent in nature, the ranking reversal phenomenon does not occur under approximate consistency.
As compared to the results of \cite{Saaty1984b}, the present finding is based on the modified axioms.
Furthermore, it should be pointed out that when a pairwise comparison matrix is not of approximate consistency, the ranking reversal phenomenon may not occur. For instance, we change the value of an entry in $A_{1}$ to get
$$
A_{2}=\left(
\begin{array}{c|ccccc}
C&x_{1}&x_{2}&x_{3}&x_{4}&x_{5}\\
\hline
x_{1}&1&9&2&3&5\\
x_{2}&1/9&1&1/5&1/3&1/2\\
x_{3}&\mathbf{1/4}&5&1&3/2&3\\
x_{4}&1/3&3&2/3&1&3/2\\
x_{5}&1/5&2&1/3&2/3&1\\
\end{array}
\right),
$$
with $a_{13}a_{31}=1/2<1.$ It is seen that $A_{2}$ is not of approximate consistency. But when eliminating an alternative, the ranking of the remaining alternatives can be kept by using the eigenvalue method. This means that the results in Theorems \ref{th8} and \ref{th9} give some ideal cases and the occurrence of ranking reversal should be further investigated.

\subsection{Multi-criteria case}

When the multiple criteria should be considered in a decision making problem, a complex and disordered state about the ranking of alternatives may occur after adding or deleting an alternative. In the single criterion case, the effects of the inner dependence of alternatives on the ranking reversal have been addressed. It can be considered that the dependence of the elements in the same level may cause the ranking reversal phenomenon. In the multi-criteria case, of much importance is to investigate the effects of the dependence of different levels in a hierarchy structure on the ranking reversal. Therefore, it is interesting to discuss the particular case where all the pairwise comparison matrices are of approximate consistency. Suppose that there are a set of criteria $\mathbb{C}=\{C_{1},
C_{2}, \cdots, C_{m}\}$ and a set of alternatives $\mathbb{X}=\{x_{1}, x_{2}, \cdots, x_{n}\}.$ The weights of criteria with respect to the goal are written as
$\bar{w}=\{w_{1}, w_{2}, \cdots, w_{m}\}$ and the weights of alternatives with respect to $C_{i}$ are expressed as $\bar{w}_{i}=\{w_{1i}, w_{2i}, \cdots, w_{ni}\}.$ For convenience, the following matrix is determined:
\begin{equation}
W=\left(
\begin{array}{c|cccc}
&w_{1}&w_{2}&\cdots&w_{m}\\
&C_{1}&C_{2}&\cdots&C_{m}\\
\hline
x_{1}&w_{11}&w_{12}&\cdots&w_{1m}\\
x_{2}&w_{21}&w_{22}&\cdots&w_{2m}\\
\vdots&\vdots&\vdots&\vdots&\vdots\\
x_{n}&w_{n1}&w_{n2}&\cdots&w_{nm}\\
\end{array}
\right).\label{eq26}
\end{equation}
Following \cite{Saaty86}, the final weights of alternatives are computed by using the following formula:
\begin{equation}
\omega_{i}=\sum_{j=1}^{m}w_{j}w_{ij},\quad i\in \mathbb{I}.\label{eq27}
\end{equation}
Without loss of generality, it is further assumed that $w_{1}\geq w_{2}\geq \cdots \geq w_{m}.$ Then we have the following results:
\begin{theorem}\label{th10}
Let the weights of criteria and alternatives be expressed as the matrix $W.$ The rankings of alternatives with respect to all criteria are assumed to be identical.
When deleting an alternative or a criterion, the final ranking of the remaining alternatives is not changed.
\end{theorem}
\proof{Proof of Theorem \ref{th10}.}
It is convenient to assume
\begin{equation}
w_{1j}\geq w_{2j}\geq \cdots\geq w_{nj},\quad \forall j\in \{1,2,\cdots, m\}.\label{eq28}
\end{equation}
Then we have $\omega_{1}\geq\omega_{2}\geq\cdots\geq \omega_{n}$ and $x_{1}\succeq x_{2}\succeq \cdots\succeq x_{n}$ by using (\ref{eq27}) and $w_{1}\geq w_{2}\geq \cdots \geq w_{m}.$
When deleting an alternative, the ranking of the remaining alternatives with respect to any a criterion is not changed according to Theorem \ref{th8}. This means that the ranking of the remaining alternatives holds by using (\ref{eq27}). In addition, when deleting a criterion, the ranking of the remaining criteria is still kept. The final ranking of the alternatives is not changed in terms of (\ref{eq27}).

\endproof
\begin{theorem}\label{th11}
The weights of criteria and alternatives are assumed to be expressed as the matrix $W,$ and there is the same ranking of alternatives with respect to all criteria.
When adding an alternative $x_{n+1},$ if the rankings of $n+1$ alternatives are identical with respect to all criteria, the ranking of the old alternatives is not changed. When adding a criterion $C_{m+1},$ if the ranking of the alternatives with respect to $C_{m+1}$ is the same as the previous one, the final ranking of alternatives is not changed.
\end{theorem}
\proof{Proof of Theorem \ref{th11}.} Following Theorem \ref{th9} and (\ref{eq27}), the results are satisfied and the detail procedure has been omitted here.

\endproof

The above observations show some ideal cases without the ranking reversal. As an example, we consider the following matrix:
$$
W_{1}=\left(
\begin{array}{c|cccc}
&0.25&0.25&0.25&0.25\\
&C_{1}&C_{2}&C_{3}&C_{4}\\
\hline
x_{1}&1/18&1/11&1/14&1/9\\
x_{2}&9/18&9/11&9/14&5/9\\
x_{3}&8/18&1/11&4/14&3/9\\
\end{array}
\right).
$$
It is seen from $W_{1}$ that $x_{2}\succeq x_{3} \succeq x_{1}$ for all criteria. If adding an alternative and giving the new matrix:
$$
W_{2}=\left(
\begin{array}{c|cccc}
&0.25&0.25&0.25&0.25\\
&C_{1}&C_{2}&C_{3}&C_{4}\\
\hline
x_{1}&1/22&1/12&1/16&1/12\\
x_{2}&9/22&9/12&9/16&5/12\\
x_{3}&8/22&1/12&4/16&3/12\\
x_{4}&4/22&1/12&2/16&2/12\\
\end{array}
\right),
$$
we can obtain the ranking of $x_{2}\succeq x_{4}\succeq x_{3} \succeq x_{1},$ meaning that the ranking of $x_{2}\succeq x_{3} \succeq x_{1}$ holds.

\subsection{Likelihood of ranking reversal}

The findings in Theorems \ref{th8}-\ref{th11} have revealed some basic facts related to the ranking reversal phenomenon. For convenience, we propose the following concept:
\begin{definition}
If the ranking of alternatives in a hierarchy structure is not changed by the actions of adding or deleting an alternative and/or a criterion, the hierarchy structure is called to be with the ranking equilibrium.
\end{definition}

The concept of ranking equilibrium describes the fundamental state of a hierarchy structure without ranking reversal. As compared to the concepts of consistency in \cite{Saaty80} and approximate consistency in Definition \ref{dj11}, the meaning of ranking equilibrium is intuitive and easily accessible. Theorems \ref{th8}-\ref{th11} can be considered as the basic cores of the ranking equilibrium.
When there is a difference among the rankings of the alternatives with respect to different row/column vectors in a pairwise comparison matrix or different criteria, the ranking equilibrium begins to be broken. In order to quantify the breaking degree of ranking equilibrium, we introduce the Kendall's coefficient of concordance \citep{kendall1970rank,field2014k}.
\begin{definition}
Assume that $\vec{V}= (v_1,v_2,\dots, v_n)^{T}$ is an $n$-dimensional vector. The rank of the element $v_{i}$ is defined as:
\begin{align}
r_{i} = \sum_{j = 1}^{ n } Ind(v_j \leq v_i),
\end{align}
where $Ind(v_j \leq v_i)$ stands for the following indicator function:
\begin{align}
Ind(v_j \leq v_i) = \begin{cases}
	1 , & 	v_j \leq v_i,\\
	0 , &  otherwise.
\end{cases}
\end{align}
Then the vector $\vec{R}(\vec{V})=(r_{1}, r_{2}, \cdots, r_{n})^{T}$ is called the rank vector of $\vec{V}.$
\end{definition}

For example, the rank vector of $\vec{V}_{1}=(0.7, 0.4, 0.3, 0.4, 0.2)^{T}$ can be determined as $\vec{R}(\vec{V}_1) = (5, 3, 2, 4, 1)^{T}.$ Hereafter, the first method is used to deal with the case with the same entries such that the rank vector of an $n$-dimensional vector is a permutation of $(1,2,\cdots,n)^{T}.$ The first method obeys the following two rules:
\begin{itemize}
  \item The rank vector should be kept as a permutation of $(1,2,\cdots,n)^{T};$
  \item For the two same entries, the rank of the entry in the first order is less than that of the second one.
\end{itemize}

In what follows, we always assume that the rank vector of an $n$-dimensional vector is a permutation of $(1,2,\cdots,n)^{T}.$
Moreover, the rank matrix can be defined as:
\begin{definition}
$R=(r_{ij})_{n \times m}$ is called the rank matrix of $B=(b_{ij})_{n\times m},$ where $r_{ij}$ is the rank of the entry $b_{ij}$ in the column vector $\vec{b}_{\cdot j}=(b_{1j}, b_{2j}, \dots, b_{nj})^{T}.$
\end{definition}

Then, let us define the following quantity:
\begin{align}\label{eq:rank-matrix}
\bar{r}_{i}=\sum_{j=1}^{m}r_{ij},\quad i\in \mathbb{I}.
\end{align}
The mean value and the variance are computed as:
\begin{align}
\bar{r}=\dfrac{1}{n}\sum_{i=1}^{n}\bar{r}_{i},  \hspace{0.5cm} S=\sum_{i=1}^{n}(\bar{r}_{i}-\bar{r})^2=\sum_{i=1}^{n}\bar{r}_{i}^{2}-n\bar{r}^{2}. \label{eq:r-S-2}
\end{align}
After some computations, we further have the following results:
\begin{align}
\bar{r}=\dfrac{n(n+1)}{2}, \hspace{0.5cm}
S=\sum_{i=1}^{n}\bar{r}_{i}^2-\dfrac{m^{2}n(n+1)^2}{4}.		
\end{align}
It is found that the maximum value of $S$ can be obtained as
\begin{align}
S_{max}=\dfrac{m^{2}n(n^2-1)}{12},
\end{align}
when the rankings of the entries of each column in the rank matrix are identical.
Now the Kendall's coefficient of concordance of $B= (b_{ij})_{n\times m}$ is defined as:
\begin{equation}
K(B)=\frac{S}{S_{max}}=\frac{12S}{m^{2}n(n^2-1)},\hspace{0.5cm} n>1.\label{eq35}
\end{equation}

Obviously, it follows $0\leq K(B)\leq1.$ When $K(B)=0,$ the column and row vectors of $R=(r_{ij})_{n \times n }$ are an identical permutation of $(1,2,\cdots,n)^{T}.$ The case of $K(B)=1$ means that the elements in a row vector of $R=(r_{ij})_{n \times n}$ are the same. In the following, the cases with single criterion and multiple criteria are investigated, respectively.

{\bf (I) The case of a single criterion}

First, let us consider the effects of the dependence of the elements in the same level on ranking reversal. That is,
when there is only a single criterion, a pairwise comparison matrix $A=(a_{ij})_{n\times n}$ should be considered.
 Under the consideration of the reciprocal symmetry breaking, the Kendall's coefficient of concordance of $A=(a_{ij})_{n\times n}$ should be derived by using the row and column vectors. The corresponding rank matrices are written as:
  $$
R_{A}^{c}=\left(
\begin{array}{c|cccc}
C&x_{1}&x_{2}&\cdots&x_{n}\\
\hline
x_{1}&r_{11}^{c}&r_{12}^{c}&\cdots&r_{1n}^{c}\\
x_{2}&r_{21}^{c}&r_{22}&\cdots&r_{2n}^{c}\\
\vdots&\vdots&\vdots&\vdots&\vdots\\
x_{n}&r_{n1}^{c}&r_{n2}^{c}&\cdots&r_{nn}^{c}\\
\end{array}
\right),
\quad
R_{A}^{r}=\left(
\begin{array}{c|cccc}
C&x_{1}&x_{2}&\cdots&x_{n}\\
\hline
x_{1}&r_{11}^{r}&r_{12}^{r}&\cdots&r_{1n}^{r}\\
x_{2}&r_{21}^{r}&r_{22}^{r}&\cdots&r_{2n}^{r}\\
\vdots&\vdots&\vdots&\vdots&\vdots\\
x_{n}&r_{n1}^{r}&r_{n2}^{r}&\cdots&r_{nn}^{r}\\
\end{array}
\right),
$$
where $R_{A}^{r}$ and $R_{A}^{c}$ stand for the rank matrices obtained by using the row and column vectors, respectively.
The Kendall's coefficient of concordance in (\ref{eq35}) should be rewritten as
\begin{equation}
K(A)=\dfrac{S^{r}+S^{c}}{2S_{max}}=\dfrac{6(S^{r}+S^{c})}{n^{3}(n^2-1)},\hspace{1cm}n>1,\label{eq36}
  \end{equation}
where
$$
S^{r}=\sum_{j=1}^{n}(\bar{r}_{j}^{r})^2-\dfrac{n^{3}(n+1)^2}{4},\quad
S^{c}=\sum_{i=1}^{n}(\bar{r}_{i}^{c})^2-\dfrac{n^{3}(n+1)^2}{4},
$$
with
$$
\bar{r}_{j}^{r}=\sum_{i=1}^{n}r_{ij}^{r},\quad \bar{r}_{i}^{c}=\sum_{j=1}^{n}r_{ij}^{c},\quad i,j\in \mathbb{I}.
$$
Here $r_{ij}^{r}$ and $r_{ij}^{c}$ are the ranks of $a_{ij}$ in the row and column vectors, respectively. We further obtain the following result:

\begin{theorem}\label{th12}
Given a pairwise comparison matrix $A=(a_{ij})_{n\times n},$ $A=(a_{ij})_{n\times n}\in\mathbb{R}_{AC(n)}$ if and only if the value of $K(A)$ in (\ref{eq36}) is equal to 1.
\end{theorem}
\proof{Proof of Theorem \ref{th12}.} When $A=(a_{ij})_{n\times n}\in\mathbb{R}_{AC(n)},$ the rankings of alternatives are identical according to all row and column vectors in $A=(a_{ij})_{n\times n}.$ This means that each column in $R_{A}^{c}$ is the same permutation of $(1,2,\cdots, n)^{T},$ and each row in $R_{A}^{r}$ is the same permutation of $(1,2,\cdots, n).$ It can be computed that $S^{r}=S^{c}=S_{max},$ then $K(A)=1.$

On the contrary, if $K(A)=1,$ it follows $S^{r}=S_{max}$ and $S^{c}=S_{max}$ are satisfied simultaneously. Then all column vectors in $R_{A}^{c}$ are identical and all row vectors in
$R_{A}^{r}$ are the same. In what follows, we prove that the same ranking of alternatives can be obtained by using each column and row vectors of $A=(a_{ij})_{n\times n}.$ The proof can be achieved by using mathematical induction.
\begin{enumerate}
  \item Let $n=2,$ meaning that
  $$
A_{(2)}=\left(
\begin{array}{c|cc}
C&x_{1}&x_{2}\\
\hline
x_{1}&1&a_{12}\\
x_{2}&a_{21}&1\\
\end{array}
\right),
$$
with $0<a_{12}a_{21}\leq1$ under Axiom 1$^\prime$. Without loss of generality, letting $a_{21}\leq1,$ it must have $a_{12}\geq1$ because the column vectors $(1, a_{21})^{T}$ and
$(a_{12}, 1)^{T}$ give the same ranking of alternatives $x_{1}\succeq x_{2}.$ This results $1\leq a_{12}$ and $a_{21}\leq1,$ implying that the application of the row vectors yields
the same ranking of $x_{1}\succeq x_{2}.$ That is, the matrix $A_{(2)}$ is of approximate consistency.
  \item It is assumed that when $n=k,$ the matrix $A_{(k)}=(a_{ij})_{k\times k}$ is of approximate consistency. Let us say $x_{1}\succeq x_{2}\succeq\cdots\succeq x_{k}$ with
  \begin{equation}
  a_{1j}\geq a_{2j}\geq\cdots\geq a_{kj},\quad a_{i1}\leq a_{i2}\leq\cdots\leq a_{ik}, \quad i,j\in\{1,2,\cdots,k\}.\label{eq37}
  \end{equation}
  \item When $n=k+1,$ the matrix $A_{(k+1)}=(a_{ij})_{(k+1)\times (k+1)}$ is given as
  $$
A_{(k+1)}=\left(
\begin{array}{c|ccccc}
C&x_{1}&x_{2}&\cdots&x_{k}&x_{k+1}\\
\hline
x_{1}&1&a_{12}&\cdots&a_{1k}&a_{1(k+1)}\\
x_{2}&a_{21}&1&\cdots&a_{2k}&a_{2(k+1)}\\
\vdots&\vdots&\vdots&\vdots&\vdots&\vdots\\
x_{k}&a_{k1}&a_{k2}&\cdots&1&a_{k(k+1)}\\
x_{k+1}&a_{(k+1)1}&a_{(k+1)2}&\cdots&a_{(k+1)k}&1\\
\end{array}
\right).
$$
By considering the same ranking of alternatives according to the column vectors of $A_{(k+1)}$
and the first relation in (\ref{eq37}), it is supposed that $a_{1j}\geq a_{2j}\geq\cdots\geq a_{kj}\geq a_{(k+1)j}$ for $j\in\{1,2,\cdots,k+1\},$
which yields the ranking of $x_{1}\succeq x_{2}\succeq\cdots\succeq x_{k}\succeq x_{k+1}.$ Then in the $(k+1)$th column of $A_{(k+1)},$ one has $a_{k(k+1)}\geq1.$
In virtue of the $k$th row of $A_{(k+1)}$ and the second equality in (\ref{eq37}), it follows $a_{k1}\leq a_{k2}\leq\cdots\leq a_{kk}\leq a_{k(k+1)}.$ Moreover, it is seen that the row vectors of $A_{(k+1)}$ should give the same
ranking of alternatives according to the rank matrix $R_{A}^{r}.$ Hence, we always have $a_{i1}\leq a_{i2}\leq\cdots\leq a_{ik}\leq a_{i(k+1)}$ for $i\in\{1,2,\cdots, k+1\},$  meaning that $x_{1}\succeq x_{2}\succeq\cdots\succeq x_{k}\succeq x_{k+1}$ by using each row of $A_{(k+1)}.$
\end{enumerate}
The above procedure shows that $A=(a_{ij})_{n\times n}\in\mathbb{R}_{AC(n)}$ under the condition of $K(A)=1.$

\endproof

Now the possibility degree of ranking reversal is defined as follows:
\begin{definition}
Let $A=(a_{ij})_{n\times n}$ be a pairwise comparison matrix. The possibility degree of ranking reversal can be quantified by using the following index:
\begin{equation}
p_{d}(A)=1-K(A).\label{eq38a}
\end{equation}
\end{definition}
When $p_{d}(A)=0$ along with $K(A)=1,$ we have $A=(a_{ij})_{n\times n}\in R_{AC(n)}.$ Based on Theorem \ref{th8} the phenomenon of ranking reversal does not occur when deleting an alternative. Under Theorem \ref{th9}, if a pairwise comparison matrix with approximate consistency is produced by adding an alternative, there is not the phenomenon of ranking reversal. The more the value of  $p_{d}(A),$ the more the likelihood of ranking reversal is. For example, let us investigate the matrix $A_{1}$ and arrive at:
$$
R_{A1}^{c}=\left(
\begin{array}{c|ccccc}
C&x_{1}&x_{2}&x_{3}&x_{4}&x_{5}\\
\hline
x_{1}&5&5&5&5&5\\
x_{2}&1&1&1&1&1\\
x_{3}&4&4&4&4&4\\
x_{4}&3&3&3&3&3\\
x_{5}&2&2&2&2&2\\
\end{array}
\right),
\quad
R_{A1}^{r}=\left(
\begin{array}{c|ccccc}
C&x_{1}&x_{2}&x_{3}&x_{4}&x_{5}\\
\hline
x_{1}&1&5&2&3&4\\
x_{2}&1&5&2&3&4\\
x_{3}&1&5&2&3&4\\
x_{4}&1&5&2&3&4\\
x_{5}&1&5&2&3&4\\
\end{array}
\right).
$$
It is easy to compute that $p_{d}(A_{1})=0$ together with $K(A_{1})=1,$ meaning that the hierarchy structure is of ranking equilibrium.
Moreover, we consider the following matrix:
$$A_{2}^{m}=\left(
\begin{array}{c|ccccc}
C&x_{1}&x_{2}&x_{3}&x_{4}&x_{5}\\
\hline
x_{1}&1&\mathbf{1/9}&2&3&5\\
x_{2}&\mathbf{9}&1&1/5&1/3&1/2\\
x_{3}&\mathbf{1/4}&5&1&3/2&3\\
x_{4}&1/3&3&2/3&1&3/2\\
x_{5}&1/5&2&1/3&2/3&1\\
\end{array}
\right),
$$
and obtain the rank matrices as:
$$
R_{A}^{c}=\left(
\begin{array}{c|ccccc}
C&x_{1}&x_{2}&x_{3}&x_{4}&x_{5}\\
\hline
x_{1}&4&1&5&5&5\\
x_{2}&5&2&1&1&1\\
x_{3}&2&5&4&4&4\\
x_{4}&3&4&3&3&3\\
x_{5}&1&3&2&2&2\\
\end{array}
\right),
\quad
R_{A}^{r}=\left(
\begin{array}{c|ccccc}
C&x_{1}&x_{2}&x_{3}&x_{4}&x_{5}\\
\hline
x_{1}&2&1&3&4&5\\
x_{2}&5&4&1&2&3\\
x_{3}&1&5&2&3&4\\
x_{4}&1&5&2&3&4\\
x_{5}&1&5&2&3&4\\
\end{array}
\right).
$$
The possibility degree of ranking reversal can be calculated as $p_{d}(A_{2}^{m})=0.6160>0.$ This means that when adding or deleting an alternative, the possibility of ranking reversal is about $61.6\%.$

{\bf (II) The case of multiple criteria}

 Second, we investigate the effects of the dependence of different levels on ranking reversal.
For the case with multiple criteria, the matrix $W$ in (\ref{eq26}) only needs to be investigated here.
Then the rank matrix can be written as:
\begin{equation}
R_{W}=\left(
\begin{array}{c|cccc}
&w_{1}&w_{2}&\cdots&w_{m}\\
&C_{1}&C_{2}&\cdots&C_{m}\\
\hline
x_{1}&r_{11}&r_{12}&\cdots&r_{1m}\\
x_{2}&r_{21}&r_{22}&\cdots&r_{2m}\\
\vdots&\vdots&\vdots&\vdots&\vdots\\
x_{n}&r_{n1}&r_{n2}&\cdots&r_{nm}\\
\end{array}
\right).\label{eq38}
\end{equation}
Similar to Theorem \ref{th12}, the following result is given:
 \begin{theorem}\label{th13}
 Assume that the weights of alternatives with respect to criteria are written as the matrix $W$ in (\ref{eq26}). The rankings of alternatives are identical according to all columns in $W$ if and only if $K(W)=1.$
 \end{theorem}
\proof{Proof of Theorem \ref{th13}.}
It is supposed that the rankings of alternatives based on all columns in $W$ are identical. It is no loss of generality to assume $x_{1}\succeq x_{2}\succeq \cdots\succeq x_{n},$
meaning that $(r_{1j}, r_{2j}, \cdots, r_{nj})=(n,n-1,\cdots,1)$ for $\forall j\in\{1,2,\cdots,m\}.$ Then we have $K(W)=1$ by using (\ref{eq35}).
On the other hand, let $K(W)=1,$ which leads to $S=S_{max}$ in (\ref{eq35}). That is, the rankings of the elements in all columns of $R_{W}$ should be identical. Hence, the rankings of alternatives determined by applying all columns in (\ref{eq26}) are the same.

\endproof

The Kendall's coefficient of concordance $K(W)$ is used to define the possibility degree of ranking reversal.
 \begin{definition}
 Assume that all pairwise comparison matrices are of approximate consistency. The derived matrix with the priorities of alternatives and criteria is shown as $W$ in (\ref{eq26}).
 The possibility degree of ranking reversal is computed as
 \begin{equation}
 p_{d}(W)=1-K(W).
 \end{equation}
 \end{definition}
It is noted that the conditions in Theorems \ref{th10} and \ref{th11} correspond to the case of $p_{d}(W)=0$ with $K(W)=1.$
The less the value of $p_{d}(W),$ the less the possibility occurring the ranking reversal phenomenon is. For example, let us consider the following matrix \citep{Dyer1990a}:
$$
W_{3}=\left(
\begin{array}{c|cccc}
&0.25&0.25&0.25&0.25\\
&C_{1}&C_{2}&C_{3}&C_{4}\\
\hline
x_{1}&1/18&9/11&1/14&3/9\\
x_{2}&9/18&1/11&9/14&1/9\\
x_{3}&8/18&1/11&4/14&5/9\\
\end{array}
\right).
$$
The corresponding rank matrix is given as:
$$
R_{W3}=\left(
\begin{array}{c|cccc}
&0.25&0.25&0.25&0.25\\
&C_{1}&C_{2}&C_{3}&C_{4}\\
\hline
x_{1}&1&3&1&2\\
x_{2}&3&1&3&1\\
x_{3}&2&2&2&3\\
\end{array}
\right).
$$
Then the Kendall's coefficient of concordance $K(W_{3})$ is calculated as $K(W_{3})=0.0625.$ That is, the possibility degree of ranking reversal is $p_{d}(W_{3})=0.9375,$
meaning that the likelihood of ranking reversal is about $93.75\%$ when adding or deleting an alternative. The above result is in accordance with the known finding to some extent \citep{Dyer1990a}.

{\bf (III) General case of a hierarchy structure}

 In general, the ranking reversal phenomenon could occur due to the dependence of the elements in the same level and different levels in a hierarchy structure. In other words, for a general case, the above two cases should be combined to consider. Then we define the possibility degree of ranking reversal as follows:
\begin{definition}\label{def17}
Let a hierarchy structure have a set of criteria $\mathbb{C}=\{C_{1},
C_{2}, \cdots, C_{m}\}$ and a set of alternatives $\mathbb{X}=\{x_{1}, x_{2}, \cdots, x_{n}\}.$ The pairwise comparisons of criteria construct the matrix $A_{C}.$ With respect to a criterion $C_{i},$ the pairwise comparisons of alternatives form the matrices $A_{C(i)}$ $(i\in \mathbb{I}).$ The relation between the levels of criteria and alternatives is expressed as the matrix $W.$ The average possibility degree of ranking reversal is computed as:
\begin{equation}
p_{d}(G)=\sum_{i=1}^{n}\nu_{i}p_{d}(A_{C(i)})+\nu_{c}p_{d}(A_{C})+\nu_{w}p_{d}(W),\label{eq40}
\end{equation}
where $\nu_{i}, \nu_{c}, \nu_{w}\in[0,1]$ for $i\in I$ and
$$\sum_{i=1}^{n}\nu_{i}+\nu_{c}+\nu_{w}=1.$$
\end{definition}

Definition \ref{def17} shows that the ranking equilibrium of a hierarchy structure is based on all pairwise comparison matrices and the structural dependence. If there is a matrix such that the ranking equilibrium is easily broken,
the ranking reversal phenomenon could occur. When all pairwise comparison matrices are of approximate consistency, the ranking equilibrium could be broken by the dependence of different levels. The formula in (\ref{eq40}) can be used to remind the decision maker notice the likelihood of ranking reversal in terms of a percentage.

\section{Comparison and discussion}

\begin{figure}[t]
\begin{center}
\includegraphics[height=1.8in]{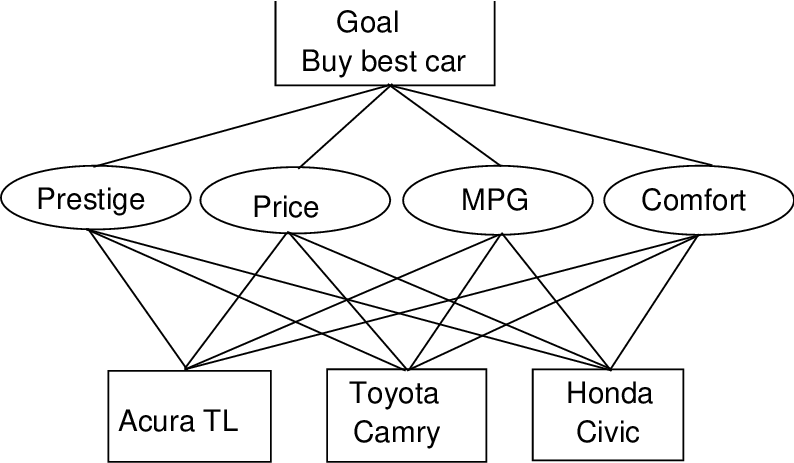}
\caption{The hierarchy structure for choosing the best car.} \label{figure3}
\end{center}
\end{figure}

For the sake of comparison, a practical case is investigated in the following. In order to choose the best car among three alternatives of Acura TL, Toyota Camry and Honda Civic, four criteria with prestige, price, miles per gallon (MPG) and comfort are taken into consideration \citep{Saaty2013}. The hierarchy structure is shown in Figure \ref{figure3} and the corresponding pairwise comparison matrices are presented in Tables 1-5, respectively. Using the eigenvalue method, the priorities of criteria and alternatives are computed. It is assumed that the final weights of three alternatives are written as $\omega_{1},$ $\omega_{2}$ and $\omega_{3},$ respectively. The obtained results are determined as
$(\omega_{1}, \omega_{2}, \omega_{3})=(0.3443, 0.2002, 0.4556).$
This means that the best car is the Honda Civic \citep{Saaty2013}.

\begin{table}[ptb]\label{lab1}
 \caption{Comparisons of criteria with respect to the goal.}
\tabcolsep 0.26in
\begin{tabular}
[c]{c|ccccc}
 \hline
 Goal&Prestige&Price&MPG&Comfort&Priorities\\
 \hline
Prestige&1&1/4&1/3&1/2&0.0987\\
Price&4&1&3&3/2&0.4250\\
MPG&3&1/3&1&1/3&0.1686\\
Comfort&2&2/3&3&1&0.3078\\
\hline
\end{tabular}
\end{table}
\begin{table}[ptb]\label{lab2}
 \caption{Comparisons of cars with respect to prestige.}
 \tabcolsep 0.21in
\begin{tabular}
[c]{c|cccc}
 \hline
 Prestige&Acura TL&Toyota Camry&Honda Civic&Priorities\\
 \hline
Acura TL&1&8&4&0.7071\\
Toyota Camry&1/8&1&1/4&0.0702\\
Honda Civic&1/4&4&1&0.2227\\
\hline
\end{tabular}
\end{table}
\begin{table}[ptb]\label{lab3}
 \caption{ Comparisons of cars with respect to Price.}
  \tabcolsep 0.21in
\begin{tabular}
[c]{c|cccc}
 \hline
 Price&Acura TL&Toyota Camry&Honda Civic&Priorities\\
 \hline
Acura TL&1&1/4&1/9&0.0633\\
Toyota Camry&4&1&1/5&0.1939\\
Honda Civic&9&5&1&0.7429\\
\hline
\end{tabular}
\end{table}
\begin{table}[ptb]\label{lab4}
 \caption{Comparisons of cars with respect to MPG.}
 \tabcolsep 0.21in
\begin{tabular}
[c]{c|cccc}
 \hline
 MPG&Acura TL&Toyota Camry&Honda Civic&Priorities\\
 \hline
Acura TL&1&2/3&1/3&0.1818\\
Toyota Camry&3/2&1&1/2&0.2727\\
Honda Civic&3&2&1&0.5455\\
\hline
\end{tabular}
\end{table}
\begin{table}[ptb]\label{lab5}
 \caption{Comparisons of cars with respect to comfort.}
  \tabcolsep 0.21in
\begin{tabular}
[c]{c|cccc}
 \hline
 Comfort&Acura TL&Toyota Camry&Honda Civic&Priorities\\
 \hline
Acura TL&1&4&7&0.7049\\
Toyota Camry&1/4&1&3&0.2109\\
Honda Civic&1/7&1/3&1&0.0841\\
\hline
\end{tabular}
\end{table}

The above solution procedure is typical with the loss of some flexibility about reciprocal property. Here the uncertainty experienced by the decision maker is considered and some comparison ratios are assumed to be with the breaking of reciprocal property. For example, the comparison ratios of criteria are changed and shown in Table 6. It is seen from Table 6 that the order of the four criteria has been rearranged. In terms of Definition \ref{dj11}, the judgements in Table 6 are not of approximate consistency. One can further find that there is a possible "mistake", since the values of $3.8, 1.5, 2.8$ and $1$ in the column are not descending. This phenomenon can be used to remind the decision maker notice the possible unreasonable behavior. The final weights of alternatives can be computed as $(\omega_{1}, \omega_{2}, \omega_{3})=(0.3417, 0.1998, 0.4585)$ and the Honda Civic is still the best choice. Here we want to point out that the uncertainty could yield the inconsistency and the breaking of reciprocal reciprocity. The feedback mechanism of reminding the decision maker could be more important than an inconsistent opinion in a practical case, since people are bound to make mistakes.

\begin{table}[ptb]\label{lab6}
 \caption{Comparisons of criteria with respect to the goal under some uncertainty.}
\tabcolsep 0.26in
\begin{tabular}
[c]{c|ccccc}
 \hline
 Goal& Price&Comfort&MPG & Prestige&Priorities\\
 \hline
Price&1 &3/2&3&{\bf 3.8}&0.4292\\
Comfort&2/3&1&3&{\bf 1.5}&0.3018\\
MPG&1/3&1/3&1&{\bf 2.8}&0.1683\\
Prestige&1/4&1/2&1/3&1 &0.1007\\
\hline
\end{tabular}
\end{table}

At the end, we analyze the likelihood of ranking reversal.
The comparison ratios in Table 6 are first considered and the rank matrices are determined as:
$$
R_{c}^{c}=\left(
\begin{array}{c|cccc}
Goal&\text{Price}&\text{Comfort}&\text{MPG}&\text{Prestige}\\
\hline
\text{Price}&4&4&3&4\\
\text{Comfort}&3&3&4&2\\
\text{MPG}&2&1&2&3\\
\text{Prestige}&1&2&1&1\\
\end{array}
\right),
$$
$$
R_{c}^{r}=\left(
\begin{array}{c|cccc}
Goal&\text{Price}&\text{Comfort}&\text{MPG}&\text{Prestige}\\
\hline
\text{Price}&1&2&3&4\\
\text{Comfort}&1&2&4&3\\
\text{MPG}&1&2&3&4\\
\text{Prestige}&1&3&2&4\\
\end{array}
\right).
$$
In terms of the formula (\ref{eq38a}), the possibility degree of ranking reversal is $p_{d}(C)=0.1062.$
Moreover, we investigate the effect of the structural dependence on ranking reversal. The rank matrix is shown as follows:
$$
R_{pc}=\left(
\begin{array}{c|cccc}
&\text{Price}&\text{Comfort}&\text{MPG}&\text{Prestige}\\
\hline
\text{Acura TL}&1&3&1&3\\
\text{Toyota Camry}&2&2&2&1\\
\text{Honda Civic}&3&1&3&2\\
\end{array}
\right).
$$
After some computations, it follows $p_{d}(W)=1-K(R_{pc})=0.9375.$
In addition, it is found that the possibility degrees of pairwise comparison matrices with respect to any criterion is $1.$ When choosing $\nu_{c}=\nu_{w}=0.5$ in the formula (\ref{eq40}), the average possibility degree of ranking reversal is computed as:
$$
p_{d}(G)=\frac{1}{2}(0.1062+0.9375)=0.5219.
$$
This implies that the possibility degree of ranking reversal is about $52.19\%$ when adding or deleting an alternative and/or a criterion.
From the above analysis, it is found that the contribution of the structural dependence to ranking reversal is much bigger than the dependence of criteria.
This means that if adding a criterion, the ranking of alternatives may not be changed. However, if adding a new car to choose, the ranking of the old alternatives could be changed.

\section{Conclusions}

The investigations of the AHP choice model have been made for over forty years since it was developed by \cite{Saaty1977,Saaty80}.
There are two important problems in addressing the theory and applications of the AHP model. One is how to derive the ratio scale in relative measurement, and the other is the phenomenon
of ranking reversal. The recent work related to the former has shown that the decision-making procedure based on the AHP model is reliable \citep{Bernasconi2010}.
For the latter, numerical examples are always carried out and a theoretical analysis is lacking except for the observation in \cite{Saaty1984b}. Here it is noticed that the reciprocal property in the axiomatic foundation of the AHP model is based on the mathematical intuition. From the viewpoints of subjective measurement and the uncertainty of human-originated information, a more flexible expression of decision information is requisite. Therefore, the concept of reciprocal symmetry breaking is proposed to allow the decision information without reciprocal property. A modified axiomatic foundation of the AHP model under uncertainty has been formed. Some interesting facts derived from the modified axioms have shown that the uncertainty experienced by the decision maker has been incorporated naturally. Moreover, the ranking reversal phenomenon has been reinvestigated according to the proposed axiomatic foundation. A novel concept of ranking equilibrium has been introduced to characterize the state without ranking reversal. The likelihood of ranking reversal has been quantified by providing an index to reflect the possibility occurring ranking reversal when adding or deleting an alternative and/or a criterion. The decision maker can be reminded by using a percentage for the ranking reversal phenomenon. The observations support the conclusion that a flexible expression of decision information with the breaking of reciprocal property could be allowed to some degree in the AHP choice model.

%
%
%


\section*{Acknowledgements}

The work was supported by the National Natural Science Foundation of China (Nos. 71871072, 71571054), 2017 Guangxi high school innovation team and outstanding scholars plan, and the Guangxi Natural Science Foundation for Distinguished Young Scholars (No. 2016GXNSFFA380004).





\end{document}